\documentclass[aps,prl,superscriptaddress,twocolumn,10pt,floatfix]{revtex4-1}
\usepackage[latin1]{inputenc}
\usepackage{amsmath}
\usepackage{amsfonts}
\usepackage{amssymb}
\usepackage{graphicx}
\usepackage{epsfig}
\usepackage{epstopdf}
\usepackage{physics}
\usepackage{mathrsfs}
\usepackage{array}
\usepackage{dsfont}
\usepackage{hyperref}
\usepackage{color}

\def \bn{\begin{align}}
\def \en{\end{align}}
\def \be{\begin{equation}}
\def \ee{\end{equation}}
\def \bea{\begin{eqnarray}}
\def \eea{\end{eqnarray}}
\def \ba{\begin{array}}
\def \ea{\end{array}}

\def \etal {{\it et al.~}}

\def \emanuele#1 {#1}

\begin{document}
	  
\title{Many-body exceptional points in colliding condensates}

\author{Mati Aharonyan}
\affiliation{Department of Physics and Center for Quantum Entanglement Science and Technology (QUEST), Bar-Ilan University, Ramat Gan 5290002, Israel}
\author{Emanuele G. Dalla Torre}
\affiliation{Department of Physics and Center for Quantum Entanglement Science and Technology (QUEST), Bar-Ilan University, Ramat Gan 5290002, Israel}

\begin{abstract}
Exceptional points describe the coalescence of the eigenmodes of a non-Hermitian matrix. When an exceptional point occurs in the unitary evolution of a many-body system, it generically leads to a dynamical instability with a finite wavevector [N. Bernier \etal, Phys. Rev. Lett. 113, 065303 (2014)]. Here, we study exceptional points in the context of the counterflow instability of colliding Bose-Einstein condensates. We show that the instability of this system is due to an exceptional point in the Bogoliubov spectrum. We further clarify the connection of this effect to the Landau criterion of superfluidity and to the scattering of classical particles. We propose an experimental set-up to directly probe this exceptional point, and demonstrate its feasibility with the aid of numerical calculations. Our work fosters the observation of exceptional points in nonequilibrium many-body quantum systems.

\end{abstract}

\maketitle

\section{Introduction}

Non-Hermitian Hamiltonians are a convenient tool to describe physical processes where the energy is not conserved, such as scattering resonances, driven-dissipative steady states, or time-dependent problems. One of the most remarkable features of non-Hermitian Hamiltonians is the existence of {\it exceptional points}, where two (or more) eigenvalues merge, and their respective eigenvectors coalesce. This  situation is forbidden in Hermitian Hamiltonians, whose eigenvectors necessarily span the entire Hilbert space. 

Exceptional points are generically isolated, because 
 they occur at specific values of the system's parameters only. \emanuele{Nevertheless, they occur frequently in many-body systems, where the wavevector can play the role of a tuning parameter.} Because many-body systems support excitations with all possible wavevectors, the existence of an exceptional point can be detected as the spontaneous formation of excitations with a specific wavelength. These excitations have a finite frequency and a finite wavevector, and offer a clear example of ``type I$_{\rm o}$'' patter formation \footnote{According to the classification of Cross and Hohenberg \cite{cross1993pattern}, patterns of type $I$ have a finite wavevector and ``o'' stands for oscillatory}. \emanuele{See, for example, Ref.~\cite{feng2017non} for a review of recent experiments of many-body exceptional points in PT-symmetric photonic systems. In the context of the unitary dynamics of many-body quantum systems, this effect was described theoretically by Ref.~\cite{Bernier2014} but, so far, has not been observed experimentally.}

Here, we show that a many-body exceptional point \emanuele{can be realized} in the collision between two  superfluids. In analogy to the Landau criterion of a single superfluid~\cite{Landau_theory}, two counterflowing superfluids become dynamically unstable when their relative velocity is larger than a critical value~\cite{Mineev1974,law2001critical,castin2014landau,castin2015vitesse,abad2015counter}. This effect has been recently observed in experiments with Bose-Fermi mixtures \cite{ferrier2014mixture} and spinor condensates \cite{kim2017critical}. The dynamical instability \emanuele{is analogous to the two-stream instability of plasma \cite{terccas2009two} and is expected to gived rise to} finite-wavevector excitations~\cite{law2001critical}.  We explain that the instability is the consequence of an exceptional point, associated with the crossing between two Bogoliubov modes. In addition, we propose an experiment targeted to \emanuele{measure the unstable Bogoliubov modes directly}.

\section{Background}

\subsection{Exceptional points in two coupled oscillators} 
One of the simplest examples of an exceptional point in a closed system is given by two coupled harmonic oscillators described by the Hamiltonian
\begin{equation}
H=\omega_1\left(\frac{p_1^2}{2}+\frac{x_1^2}{2}\right)+\omega_2\left(\frac{p_2^2}{2}+\frac{x_2^2}{2}\right)+\alpha x_1 x_2,\label{eq:H12}
\end{equation}
where $x_{1/2}$ and $p_{1/2}$ are pairs of canonically conjugated variables. Although Eq.~(\ref{eq:H12}) involves only real functions of the operators $x$ and $p$, this Hamiltonian is not necessary Hermitian because, as we will see, in certain cases it is not bounded from below. By solving the equations of motion associated with the Hamiltonian of Eq.~(\ref{eq:H12}), one finds that the eigenfrequencies of this system are \cite{rossignoli2005complex,Bernier2014}
\begin{equation}
\omega^2=\frac{\omega_1^2+\omega_2^2}{2}\pm\sqrt{\left(\frac{\omega_1^2-\omega_2^2}{2}\right)^2+\alpha\omega_1\omega_2}.\label{eq:coupled}
\end{equation}
For $\alpha=0$ one simply obtains the eigenfrequencies $\omega=\pm \omega_1$ and $\omega=\pm \omega_2$. As a consequence, one has a level crossing when either $\omega_1=\omega_2$ or $\omega_1=-\omega_2$.

\begin{figure}
(a) ${\rm sign}[\omega_1]={\rm sign}[\omega_2]$ \\
\includegraphics[scale=0.4]{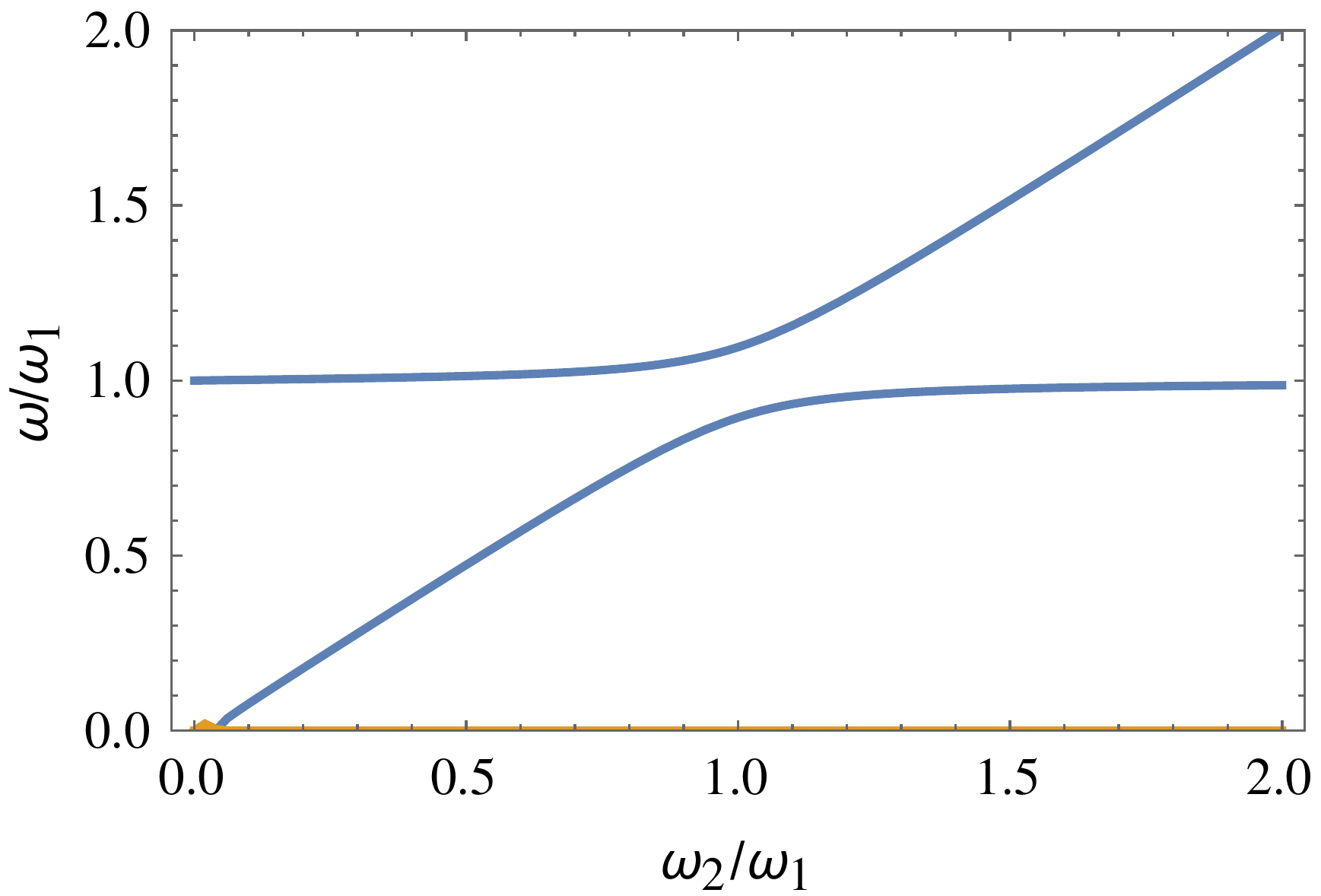}\\
(b) ${\rm sign}[\omega_1]=-{\rm sign}[\omega_2]$ \\
\includegraphics[scale=0.4]{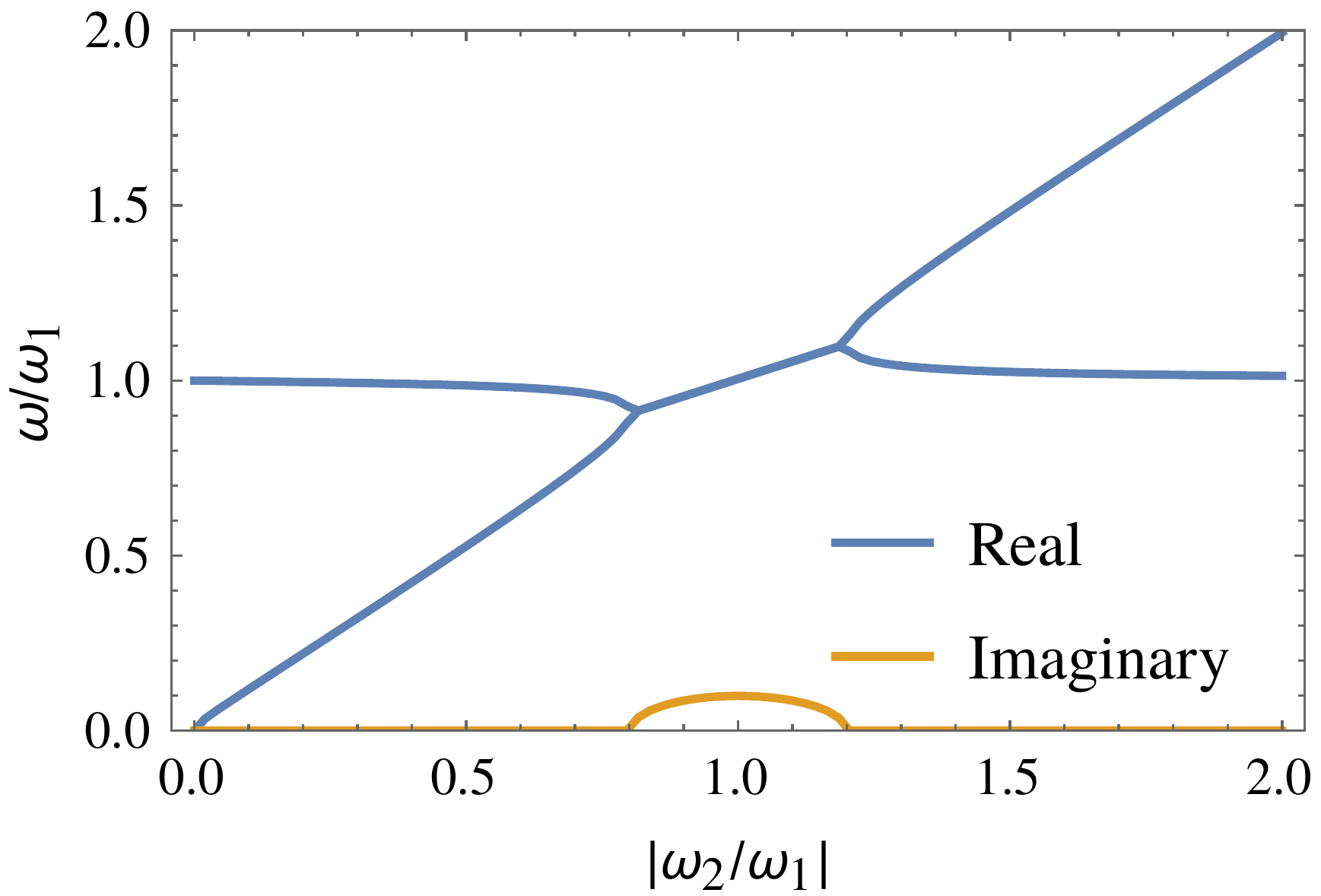}
\caption{Eigenfrequencies of a system of two coupled harmonic oscillators, Eq.~(\ref{eq:coupled}), when $\omega_1$ and $\omega_2$ have (a) the same sign, or (b) opposite signs. The former case gives rise to an avoided level crossing, while the latter leads to a dynamical instability.}
\label{fig1}
\end{figure}

The effect of a finite $\alpha$ depends on the relative sign between $\omega_1$ and $\omega_2$ \cite{skryabin2000instabilities,nakamura2008condition,Bernier2014}. If $\omega_1$ and $\omega_2$ have the same sign (or ``signature''), the two eigenfrequencies repel each other, giving rise to an avoided level crossing (Fig. \ref{fig1}(a)). 
In contrast, if $\omega_1$ and $\omega_2$ have opposite signs, when the two eigenvalues approach, the argument of the square-root of Eq.~(\ref{eq:coupled}) can vanish. This situation gives rise to an exceptional point, where two eigenmodes coalesce. In the region where the argument of the square-root is negative, the eigenfrequencies acquire a finite imaginary component, giving rise to a dynamical instability \cite{Bernier2014} (Fig. \ref{fig1}(b)). In what follows, we will show that this effect is at the origin of the counterflow instability between two condensates.

\subsection{The counterflow instability} The problem of two superfluids that flow in opposite directions was first considered theoretically in 1974 by Mineev \cite{Mineev1974}$,$\footnote{Interestingly, at that time, only one material, He$^4$ was known to Bose condense. Ref.~
\cite{Mineev1974} indeed mentions: ``Let  us  note  at  once,  however,  that  there  apparently  does  not  exist  on  this  planet  a  solution  of  two  Bose  liquids  such  that  both  components  do  not  solidify  before  the  $\lambda$  transition.  The  only  other  candidate  besides  He$^4$,  namely  He$^6$,  is  radioactive  (half-life  0.8 sec).  But  besides  possible  astrophysical  applications,  such  a  simple  microscopic  model  is  of  interest,  since  it  may  help  us  understand  the  general  laws  typical  of  solutions  of  two  superfluid  liquids. For  example,  the  presence  of  two  Bose  branches  in  the  spectrum  of  the  elementary  excitations  is  typical  of  both  Bose--Bose  and  Fermi--Bose  superfluid  mixtures.''. It is quite remarkable that more than 40 years later, this theoretical work has become relevant to actual experiments on this planet.}. By considering the ``center of mass'' and ``relative'' modes of the two superfluids, Mineev found two distinct Bogoliubov modes, which he associated with the first and third sounds, respectively \footnote{These modes have been recently probed experimentally by Ref.~\cite{fava2018observation}.}. Mineev predicted the system to remain superfluid as long as both modes are energetically stable, i.e. their sound velocities are positive. In the limit of small interspecies interactions $g_{12}\to0$, one obtains that the critical velocity equals to the smaller of the two sound velocities of the original condensates, $v_c={\rm min}[c_1,c_2]$. 

The result by Mineev refers to an impurity moving inside the counterflowing superfluids. A distinct question is whether the system can generate friction even without any impurity. Can  one condensate act as a perturbation for the other, and lead to a dynamical instability? 

This question was analyzed by Law \etal~\cite{law2001critical}, who found that counterflowing superfluids become dynamically unstable when the relative velocity exceeds a critical velocity $v_c$.  In the limit of weak interspecies interactions, $g_{12}\to0$, the critical velocity equals to the sum of the sound velocities of the two superfluids, $v_c=c_1+c_2$ \cite{abad2015counter}. The critical velocity is a monotonously decreasing function of the interspecies interaction and vanishes at $g_{12}=\sqrt{g_1g_2}$, where $g_1$ and $g_2$ are the intraspecies interactions \cite{law2001critical}. This point corresponds to the miscible-to-immiscible phase transition, where, indeed, the system becomes unstable even at zero relative velocity \cite{pu1998properties,ao1998binary,timmermans1998phase,papp2008tunable,lee2016phase},~\footnote{Conuterflowing superfluids were studied both theoretically and experimentally in earlier works. See in particular theoretical studies concerning the counterflow instability in Mott insulators \cite{Mott_counterflow}, normal fluids \cite{haber2016instabilities}, and in higher dimensions, where mean-field applies \cite{chevy2015counterflow}. In two dimensions, numerical simulations reveal that the excitations have the form of vortex-antivortex pairs, and eventually lead to a chaotic motion of the particles \cite{shungo_short,Shungo,fujimoto2012counterflow}. See also the famous quantum Newton's cradle experiment for the case of the counterflow between two condensates made of identical atoms \cite{newton_cradle}, and experiments of atomic-pair generation in colliding condensates \cite{perrin2007observation,kheruntsyan2012violation}.}.

The existence of a critical counterflow velocity has been recently observed experimentally by two groups, using different systems. The first realization involved a Bose-Fermi mixture \cite{ferrier2014mixture},~\footnote{See Refs.~\cite{castin2014landau,castin2015vitesse} for the relation between this expression and the Landau criterion of a Fermi superfluid.}: one species was positioned in the middle of the trap, while the other was initially displaced, and performed periodic oscillations at the trap frequency. By measuring the decay of the oscillations, it was found that a significant dumping occurs only when the relative velocity is larger that a critical value. This experiment was then extended by Ref.~\cite{delehaye2015critical}, who used a Feshbach resonance to enhance the interspecies interaction \cite{FESHBACH}, and observed a reduction of the critical velocity. The second realization used two hyperfine states of a single condensate \cite{kim2017critical}. These authors prepared a mixture of the two species, and induced a relative velocity through a gradient of the magnetic field. It was found that the counterflow becomes dissipative only for magnetic field larger than a critical value. 

\begin{figure}
\includegraphics{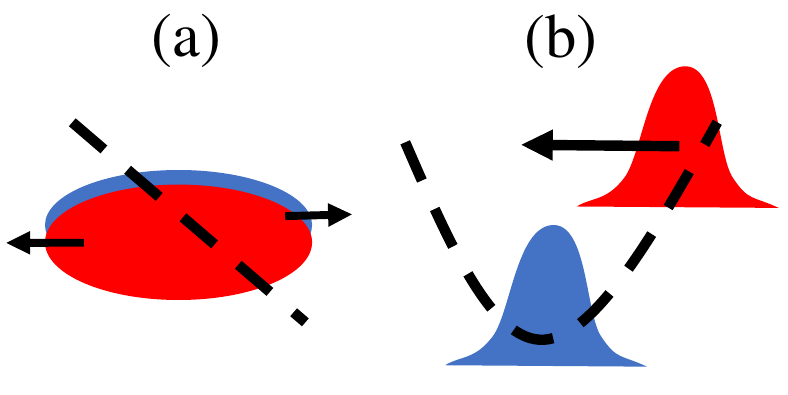}
\caption{Schematic plot of two experimental realizations of the counter-flow instability: (a) The atoms of a BEC are prepared in a superposition of two hyperfine states, and a gradient of the magnetic field is applied. The two states are accelerated in opposite directions \cite{hamner2011generation,hoefer2011dark,kim2017critical}; (b) Two species of atoms are cooled independently, and placed in different positions of an harmonic trap \cite{ferrier2014mixture}. In this work, we focus on the latter case, where the collision occurs with a constant relative velocity.}
\label{fig:schematic}
\end{figure}

A related question is what happens when the relative velocity is larger than its critical value. Law \etal \cite{law2001critical} predicted the existence of a dynamical instability at a finite wavevector. Experiments with spinor condensates showed that the system develops isolated dark-bright solitons~\cite{hamner2011generation}, or trains of dark-dark solitons~\cite{hoefer2011dark}. These solitons can be understood in terms of the modulation instability of the non linear equations of motion of the condensates \footnote{See Ref.~\cite{zakharov2009modulation} for an introduction},\cite{shukla2001modulational}. A quantitative comparison between the theoretical predictions of Law \etal and the experimental observations of Refs.~\cite{hamner2011generation,hoefer2011dark} was not possible for two main reasons: First, the theoretical study referred to a constant relative velocity, while the experiments were performed under a constant relative acceleration. Second, the solitons were observed in real space, while the theoretically predicted instabilities are characterized by a finite wavevector. In this work, we bridge this conceptual gap, by proposing an experiment in which finite-wavevector instabilities can be observed. The key difference from the experiments of Refs.~\cite{hamner2011generation,hoefer2011dark} is that we consider two condensates that are initially displaced, so that they collide with an approximately constant relative velocity (see Fig.~\ref{fig:schematic}).

\section{Methods}
\subsection{Gross-Pitaevskii and Bogoliubov modes}
Following Refs.~\cite{Mineev1974,law2001critical,abad2015counter}, we describe the counterflow of two superfluids using the one-dimensional Gross-Pitaevskii equation (GPE)  \footnote{See Ref.~\cite{stringari_review} for an introduction}
\begin{align}
\pdv[]{{\psi}_1}{t}&= i \big ( \frac{\hbar}{2m}\pdv[2]{}{x}+ g_1\abs{\psi_1}^2 +g_{12}\abs{\psi_2}^2 +V(x)\big ) {\psi}_1,  \label{gp1}\\
\pdv[]{{\psi}_2}{t}&= i \big ( \frac{\hbar}{2m}\pdv[2]{}{x}+ g_2\abs{\psi_2}^2 +g_{12}\abs{\psi_1}^2 +V(x) \big ) {\psi}_2. \label{gp2}
\end{align}
Here $V(x)$ is the confining potential, $g_{1}$, $g_2$ and $g_{12}$ are the intraspecies and interspecies coupling constants. 

In the case of a translationally invariant systems ($V(x)\equiv V_0$), the eigenfrequencies of Eqs.~(\ref{gp1}) and (\ref{gp2}) can be found analytically through the Bogoliubov transformation \cite{Mineev1974,law2001critical,abad2015counter}. For completeness, we summarize the main steps of the derivation. First, one looks for the mean-field solution of Eqs.~(\ref{gp1}) and (\ref{gp2}), which is given by
$ \psi^{(\alpha)}(x,t) = \psi_0^{(\alpha)} e^{-i\mu_1 t} $ where $\alpha=1,~2$, $\mu_1=-g_1n_1-g_{12}n_2-V_0$ and $\mu_2=-g_2n_1-g_{12}n_2-V_0$, and $n_\alpha=|\psi_0^{(\alpha)}|^2$ is the average density of particles in the $\alpha^{\rm th}$ condensate. Next, one considers small perturbations around the mean-field solution:
\begin{align}
\delta\psi^{(\alpha)}(x,t) =\sum_{k}\Big[u_k^\alpha e^{i(kx-(\mu_\alpha+\omega)t)} +v_k^\alpha e^{-i(kx+(\mu_\alpha-\omega)t)}\Big], \label{excited_psi1}
\end{align}
where $\alpha=1,~2$. Then, one assumes $u_k$ and $v_k$ to be much smaller than $\psi_0$ and linearizes Eqs. (\ref{gp1}) and (\ref{gp2}) to obtain
\begin{equation}
 \begin{pmatrix}  M_{11}-\omega \mathds{1} & M_{12}\\ M_{21} & M_{22}  -\omega \mathds{1}\end{pmatrix} \begin{pmatrix}u_k^1 \\ \bar{v}_k^{1}\\ u_k^2\\  \bar{v}_k^{2} \end{pmatrix} = 0, \label{matrix}
\end{equation}
where $\bar{\cdot}$ denotes complex conjugation and
\begin{align}
& M_{11}=\begin{pmatrix}
 \frac{\hbar k^2}{2m_1}+ {g_1|\psi_0^{(1)}|^2} & {g_1\psi_0^{(1)}}^2 &\\
 -{{g_1\bar\psi_0^{(1)}}^2} & -\frac{\hbar k^2}{2m_1} -{g_1|\psi_0^{(1)}|^2} \label{upper_left}
\end{pmatrix},\\
& M_{12}=\begin{pmatrix}
g_{12}\psi_0^{(1)}\bar\psi_0^{(2)} & g_{12}\psi_0^{(1)}\psi_0^{(2)}\\
-g_{12}\bar\psi_0^{(1)}\bar\psi_0^{(2)} & -g_{12}\bar\psi_0^{(1)}\psi_0^{(2)}
\end{pmatrix},
\\
& M_{21}= \begin{pmatrix}
g_{12}\bar\psi_0^{(1)}\psi_0^{(2)} & g_{12}\psi_0^{(1)}\bar\psi_0^{(2)}\\
-g_{12}\bar\psi_0^{(1)}\bar\psi_0^{(2)} & -g_{12}\psi_0^{(1)}\bar\psi_0^{(2)}
\end{pmatrix},\\
& M_{22}= \begin{pmatrix}
\frac{\hbar k^2}{2m_2}+ {g_2|\psi_0^{(2)}|}^2 & {g_2\psi_0^{(2)}}^2\\
-{g_2\bar\psi_0^{(2)}}^2 & -\frac{\hbar k^2}{2m_2}- {g_2|\psi_0^{(2)}|}^2 \label{bottom_right}
\end{pmatrix}.
\end{align}
Finally, the Bogoliubov spectrum can be found by demanding the matrix in Eq.~(\ref{matrix}) to be singular (i.e. its determinant to be zero).

For simplicity, in this article we consider the symmetric case $m_1=m_2=m$, $g_1=g_2=g$, and $\psi_1=\sqrt{n_1}=\psi_2=\sqrt{n_2}$. These assumptions are relevant to the case where the two species correspond to two hyperfine states of the same atoms \cite{2_specie_prperties}. In this case, the eigenfrequencies are explicitly given by:
\begin{align}
\omega^\pm_{1/2}&=\pm\sqrt{\frac{\hbar^2k^4}{4m^2}+\frac{2g_1\hbar k^2 n}{2m}\pm 2\sqrt{\hbar^2k^4n^2g_{12}^2}{4m^2}}
\label{omega4}
\end{align}
The extension to the asymmetric case is straightforward, although the analytical expressions become more cumbersome \cite{kourakis2005modulational}. Note that the positive ($\omega^+>0$) and negative ($\omega^-<0$) branches represent different physical Bogoliubov modes, and are respectively associated with particle-like and hole-like excitations \footnote{This distinction is due to the fact that the GPE involves complex wavefunctions. In contrast, in the case of a classical modes, such as elastic waves, the positive and negative branches are physically identical}. As we will now explain, the crossing between a particle-like and a hole-like excitation is at the origin of the counterflow instability.

A relative velocity between the two condensates 
can be studied by modifying the wavefunction of the first condensate according to the Galilean transformation $x\rightarrow x-v_{\rm rel}t$ in  Eq.~(\ref{excited_psi1}). Following the same procedure as before, one finds that the Galilean 
transformation adds $-v_{\rm rel}k$ to the diagonals of Eq.(\ref{upper_left})
\begin{equation}
M_{11} \to M_{11} -v_{\rm rel}k\left(\begin{array}{c c}1 & 0\\0 &1 \end{array}\right)\;.
\end{equation}
The eigenfrequencies can then be found analytically as before \cite{law2001critical,abad2015counter}.
\begin{align}
\omega_{1/2}^\pm& = -\frac12kv_{\rm rel}\pm \left[\frac{\hbar^2k^4}{m^2}+\frac{4g_1 n\hbar k^2 }m+k^2v_{\rm rel}^2\right.\nonumber\\&\left.\pm4\sqrt{\frac{\hbar^2k^6v^2}{4m^2}+\frac{g_1n \hbar k^4v^2}{m}+\frac{g_{12}^2 n^2\hbar^2k^4 }{m^2}}\right]^{1/2}
\label{eq:omega12v}
\end{align}

\begin{table}
\begin{tabular}{|l|l |l| l|}
\hline
Constants & m & Rb atomic mass & $1.44E{-25}~{\rm kg}$\\
\hline
& $g_1=g_2$ & effective 1d coupling & $8.4E{-6}$~m/s\\
\hline 
& $\hbar$ & Planck's constant & $1.054E{-34}~J s$\\
\hline
\hline
Bogoliubov & $n_1=n_2$ & density & 1000~$\mu$m$^{-1}$\\
\hline
& $c_1=c_2$ & sound velocity  & 0.0025~m/s\\ 
\hline
\hline
GPE & $N_1=N_2$ & number of particles & 5000\\
\hline
& $\omega_0$ & trap frequency & 628 rad/sec\\
\hline
& $a$ & numerical grid & 0.058 ${\mu}$m\\
\hline
\end{tabular}
\caption{Physical parameters used in the numerical calculations, and relevant for typical atom-chip experiments.\label{mytable}}
\end{table}

\begin{figure}
\centering
(a) t=0 \\
\includegraphics[scale=0.7]{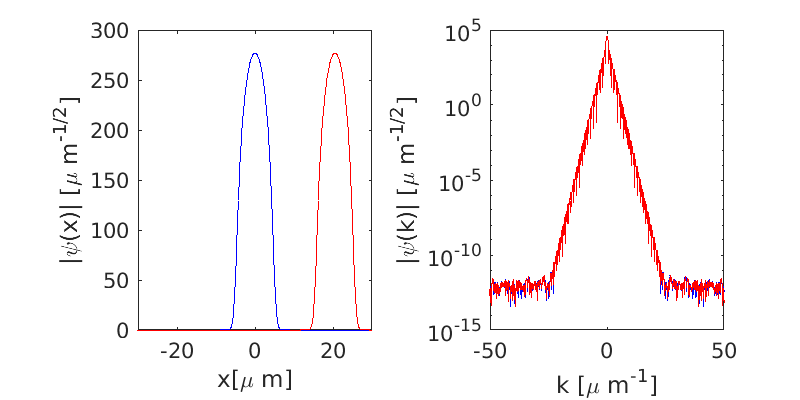}\\
(b) t=2.6ms\\
\includegraphics[scale=0.7]{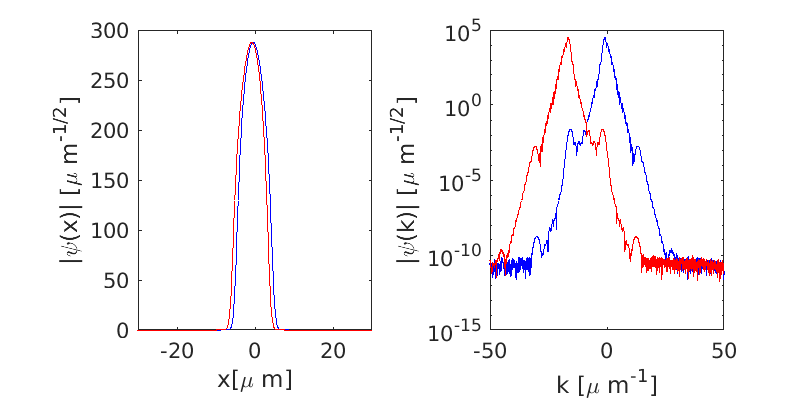}\\
(c)t=5ms\\
\includegraphics[scale=0.7]{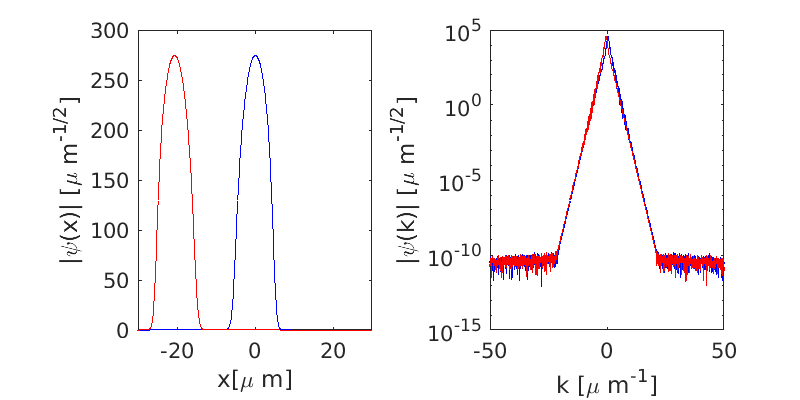}\\
(d)t=7.7ms\\
\includegraphics[scale=0.7]{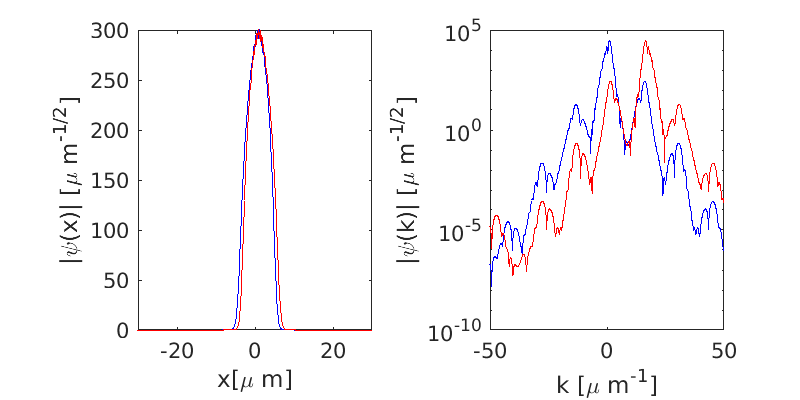} 
\caption{Time evolution of two colliding condensates in real and Fourier space for interspecies interactions $g_{12}=2g_1$. The two condensates are initially displaced by $20\mu$m, leading to a relative velocity $v_{\rm rel}\approx 0.013$m/s. All other numerical values are given in Table \ref{mytable}.}\label{collisions}
\end{figure}

\subsection{Numerical solution of the GPE}

Ultra-cold atoms are usually confined in parabolic traps. In this case, the momentum is not a good quantum number, and we are unable to solve the problem analytically. As we now explain, it is, nevertheless, still possible to draw a direct comparison with the translational invariant case. 

To describe the effects of an harmonic confinement $V(x)=(1/2)m\omega_0^2x^2$, we solved numerically the GPE, Eqs.~(\ref{gp1}) and (\ref{gp2}), following the common procedure: (i) We computed the wavefunction of each condensate using the Thomas-Fermi approximation. (ii) To improve the description of the condensates' wavefunctions, we evolved the GPE in the absence of interspecies interactions. (iii) We shifted the position of one condensate, while leaving the second condensate in the middle of trap, in analogy to the experimental situation of Refs.~\cite{maddaloni2000collective,ferrier2014mixture,delehaye2015critical}. (iv) We numerically solved the GPE in the presence of interspecies interactions, and found the time evolution of the condensates' wavefunctions. The physical parameters used in our calculations are summarized in Table \ref{mytable} and correspond to typical values for atom-chip experiments \footnote{See Ref.~\cite{keil2016fifteen} for a recent review of these experiments.}.

Fig.~\ref{collisions} shows the results of our numerical solution of the GPE for a specific combination of the initial displacement and the interspecies interaction. The left panels show the absolute value of the wavefunction: The blue condensate is at rest in the center of the trap ($x=0$), while the red oscillates in time. Each subfigure represents a different step in the time-evolution: (a) at the initial conditions, (b) during the first collision, (c) after the first collision, and (d) during the second collision.  \footnote{To facilitate the analysis of the dynamics, movies were also created and are available at http://nonequilibrium.ph.biu.ac.il/thesis/.}.

In the real-space pictures, the excitations are hardly visible due to their weak intensity with respect to the condensate. To identify the excitations and their wavevectors, we computed the Fourier transform of the wavefunctions. This quantity can be directly probed in experiments using the time-of-flight technique. The Fourier transformed wavefunctions are plotted in the right panels of Fig.~\ref{collisions}: (a) The two wavefunctions are on top of each other, because both clouds are initially at rest. (b) The blue condensate is still at rest, while the red condensate is moving to the left. In this plot, each condensate has two peaks: a main peak that describes the condensates, and two minor peaks that describe a finite-wavevector instability. (c) The two condensates overlap again, because the red condensates has reached the maximal distance from the center and is now at rest. (d) The condensates collide for a second time. At this time, the excitations are again visible, with a much larger contrast. Generically, we observe that with an increasing number of collisions, the excitations become stronger, and additional peaks at higher wavevectors are generated.


We repeated this calculation for several values of the initial displacement and of the interspecies interactions. For each run, we first determined the time of the collision and the relative velocity of the two condensates. To achieve this goal, we computed numerically the position of the center of mass of the moving condensate as a function of time and found the crossings with the origin. Next, we extracted the wavevector and the intensity of the excitations, by inspecting the wavefunction of the static condensate (blue) at the time of the first collision. This procedure was performed by subtracting the initial wavefunction (Fig.~\ref{collisions}(a)) from the solution of the GPE at the collision time (Fig.~\ref{collisions}(b)). The position and height of the maximum of this function were denoted by $k^*$ and $\delta\psi$, respectively, and are plotted in Figs.~\ref{fig:kmax}(b) and \ref{fig:Imax}(b).

In our calculations, we limited ourself to the initial configurations for which the condensates are spatially separated. For the physical values considered in this simulation, this requirement implies that the initial displacements should be larger than $\Delta x=10\mu$m. This condition restricts the values of the relative velocities that can be studied. The relation between the initial displacement and the relative velocity can be estimated using the energy conservation: $(1/2)\omega_0 (\Delta x)^2 = (1/2)mv^2$, or $v=\omega_0 \Delta x$. Our calculations were performed for $\omega_0=628 rad/sec$, leading to a minimal relative  velocity of $v_{\rm min}\approx 0.006$m/s. Note that the the sound velocities of the condensates at rest is $c_1=c_2=0.0025$m/s. Thus, our GPE calculations refer to the unstable regime only, $v_{\rm rel}>c_1+c_2>v_c$.

\section{Results}
\subsection{Criteria for the unstable level crossing}
\begin{figure}
(a) $v_{\rm rel}=0$ \\
\includegraphics[scale=0.6]{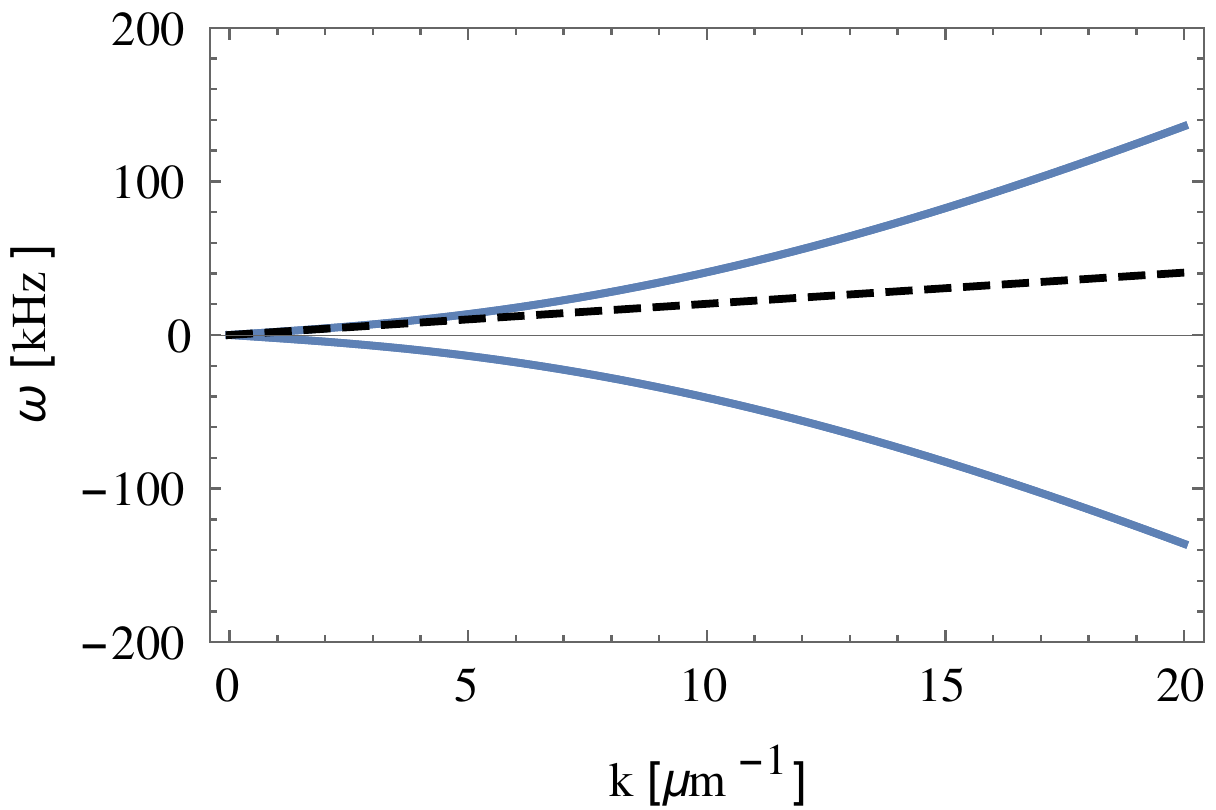} \\
(b) $v_{\rm rel}=0.01$m/s\\
\includegraphics[scale=0.6]{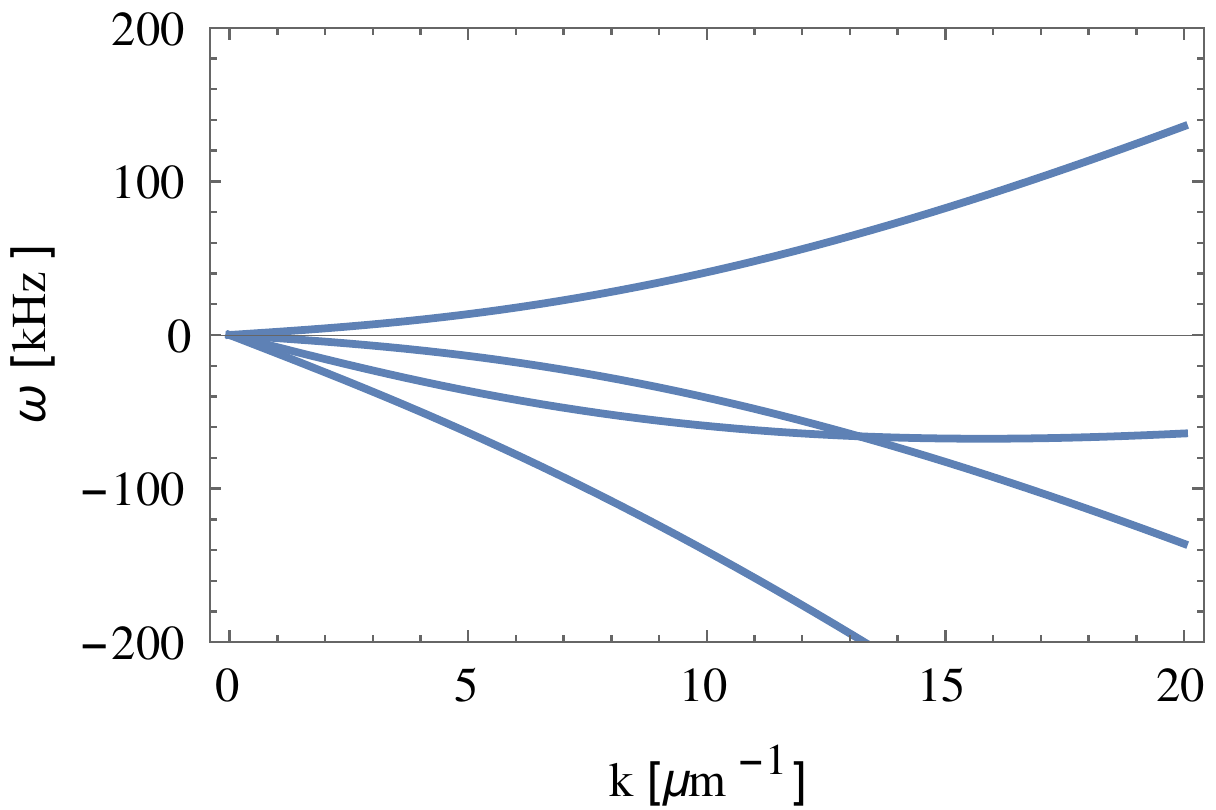}
\caption{Eigenfrequencies of the Bogoliubov modes in the limit of vanishing interspecies interactions ($g_{12}=0$) (a) at rest and (b) for a finite relative velocity. In the latter case, a crossing between Bogoliubov modes is observed. The dashed line is the linear dispersion $\omega=ck$, valid for small wavevectors. The physical parameters are given in Table~\ref{mytable}.}
\label{modes_relative_velocity}
\end{figure}

\begin{figure}
(a) $g_{12}/g_1=0$ \\
\includegraphics[scale=0.6]{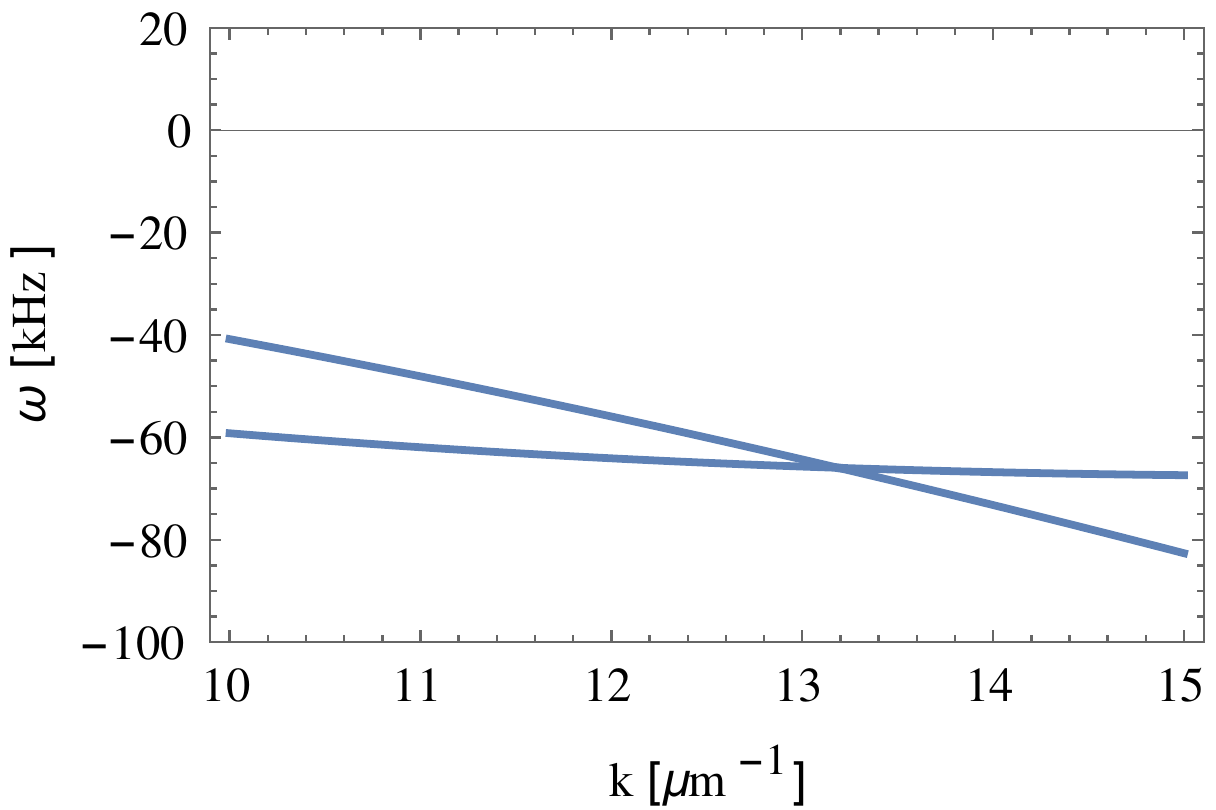} \\
(b) $g_{12}=g_{1}/2$\\
\includegraphics[scale=0.6]{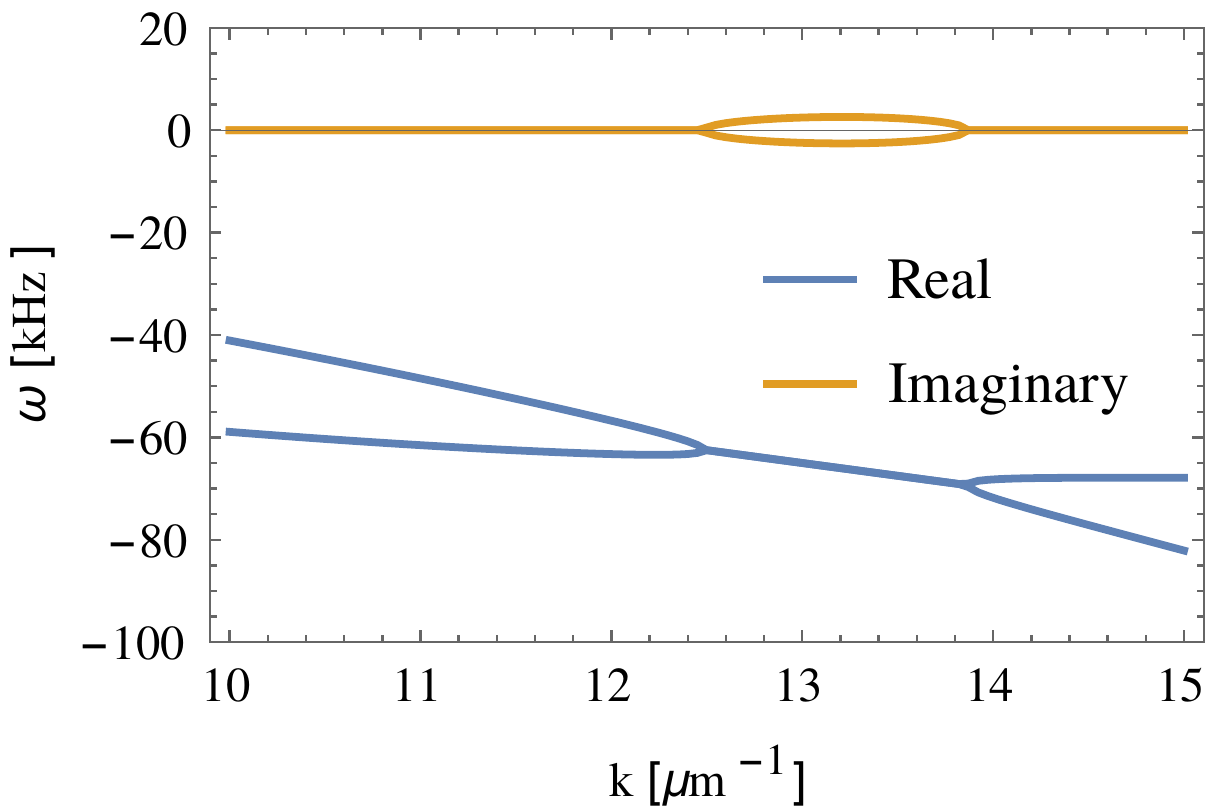}
\caption{Eigenfrequencies of the Bogoliubov modes Bogoliubov modes for a finite relative velocity $v_{\rm rel}=0.01~$m/s (a) in the absence of interspecies interactions ($g_{12}=0$) and in the presence of weak interspecies interactions ($g_{12}=g_1/2$). In the latter case, the system develops a dynamical instability at a finite wavevector.}
\label{excitations} 
\end{figure}

We now use the Bogoliubov modes to analyze the counterflow instability. Let us first consider the case of a vanishing interspecies coupling $g_{12}=0$. 
In this case, the modes of the system, Eq.~(\ref{eq:omega12v}), are simply given by $\omega^\pm_1 = \pm \omega_0 - v_{\rm rel} k$, and $\omega^\pm_2=\pm\omega_0$, where 
\begin{align}
\omega_0= \sqrt{\frac{k^4\hbar^2}{4m}+\frac{k^2\hbar g n}{m}}.\label{omega_0}
\end{align}
is the Bogoliubov spectrum of the condensate at rest. As shown in Fig.~\ref{modes_relative_velocity}, if the relative velocity is large enough, one observes a crossing between a particle like mode ($\omega_1^+$), and a hole like mode ($\omega_2^-$). When this happens, a finite interspecies interaction $g_{12}$ is sufficient to induce an unstable level crossing (see Fig.~\ref{excitations}). In the region around the crossing, the eigenfrequencies acquire a finite imaginary component, indicating a dynamical instability. This effect is at the origin of the finite-wavelength instability discussed by Law \etal~\cite{law2001critical}.

To determine the conditions for this crossing, we now introduce a linear approximation to the dispersion relation, which is valid for small wavevectors: $\omega_0 \approx c|k|$, with $c=\sqrt{g n \hbar /m}$. Under this approximation, the eigenfrequencies of the counterflowing modes are simply $\omega^\pm_{1}=\pm (c+v_{\rm rel})|k|$ and $\omega^\pm_{2}=\pm c |k|$. Two modes cross when $\omega_1^+=\omega_2^-$, or equivalently $c+v_c=-c \Rightarrow v_c=2c$. This condition corresponds to the Landau criterion for two superfluids found by Refs.~\cite{law2001critical,abad2015counter}. For finite interspecies interactions, the instability can occur for smaller relative velocities, leading to a reduction of the critical velocity. 

Let us now consider the opposite limit, of large wavevectors. In this regime, the Bogoliubov eigenfrequencies, Eq.~(\ref{omega_0}), correspond to the kinetic energy of free particles ($\omega_0\approx\hbar k^2/2m$). After the Galilean transformation, one obtains $\omega^\pm_{1}=\pm\hbar k^2 /2m + v_{\rm rel} k$ and $\omega^\pm_{2}=\pm\hbar^2k^2/2m$. The condition for an intersection between two modes becomes $\omega_1^+=\omega_2^-$, or $k=m v_{\rm rel}/\hbar$. This condition can be understood as the elastic collision between classical particles in one dimension: When an atom from a moving condensate collides with an atom from the static condensate, they exchange their velocity, effectively creating a Bogoliubov excitation at momentum $\hbar k=mv_{\rm rel}$. This finding shows that the counterflow instability of a quantum gas is adiabatically connected to the collision of classical particles.

\subsection{Finite-wavevector instability in a trap}
In this section we discuss the detection of the finite wave-vector instability in a trap, by comparing the results of the Bogoliubov and GPE methods. The quantitative agreement between these two approaches supports the feasibility of the proposed experiment. 

\begin{figure}[t]
\centering
(a) Bogoliubov \\
\includegraphics[scale=0.8]{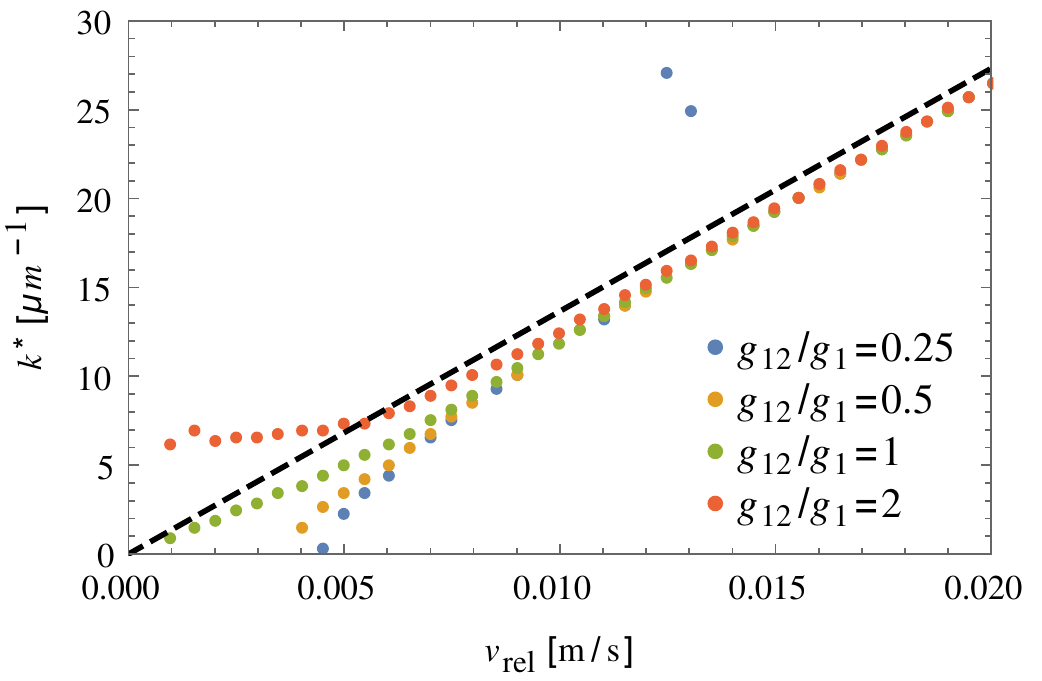}\\
(b) GPE\\
\includegraphics[scale=0.8]{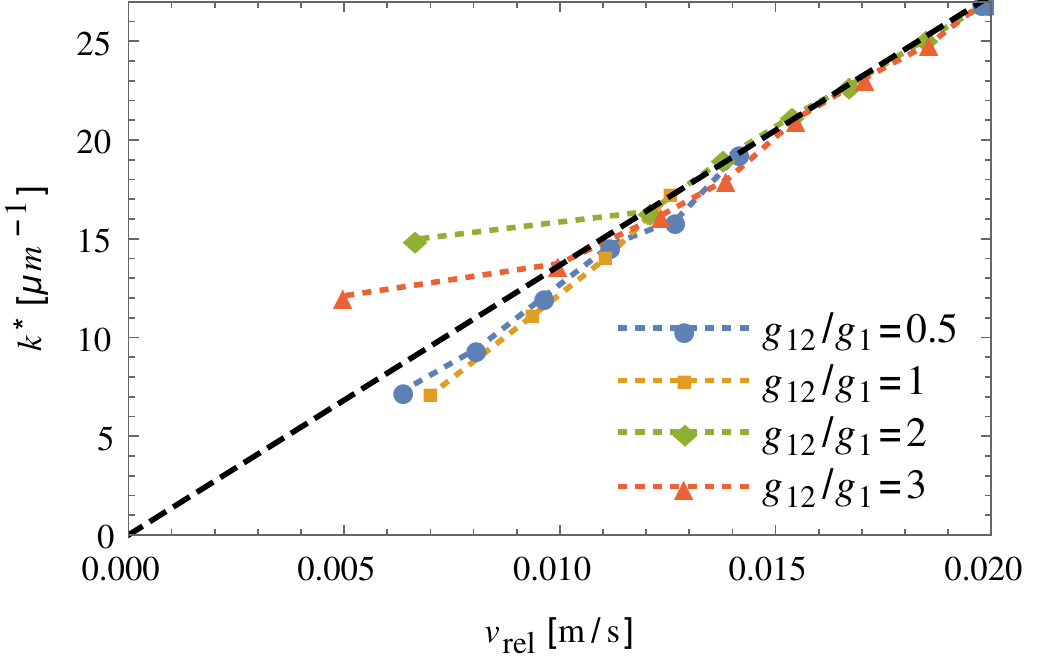}\\
\caption{Wavevector of the dominant excitation as a function of the relative velocity between the condensates, according to (a) the analytical expression of the Bogoliubov eigenfrequencies in a translationally invariant system, Eq.~(\ref{eq:omega12v}), and (b) the numerical solution of the GPE in a trap. For large relative velocities the wavevector is given by the classical expression $\hbar k = m v_{\rm rel}$ (dashed line). For small relative velocities and $g_{12}<g_1$, the excitations are suppressed and disappear below the critical velocity $v_c<2c\approx 0.005$m/s.}\label{fig:kmax}
\end{figure}

Let us first consider the excitation's wavevector as a function of the relative velocity. Fig.~\ref{fig:kmax} shows the results of (a) the Bogoliubov analysis, and (b) the GPE. In both plots the excitation wavevector is a monotonously increasing function of the relative velocity. For large relative velocities, the wavevectors does not depend on the interspecies interaction, and is approximately given by the semi-classical scattering of free particles with a quadratic dispersion, $\hbar k = m v_{\rm rel}$. For small relative velocities the behavior differs in the miscible ($g_{12}<g_1$) and immiscible ($g_{12}>g_1$) phases. In the miscible phase, the wavevector is suppressed and vanishes at the critical velocity, $v_c < 2c = 0.005 m/s$. In the immiscible phase, the system is unstable for any relative velocity, and the wavevector of the dominant excitation saturates to a finite value.


\begin{figure}[t]
\centering
(a) Bogoliubov\\
\includegraphics[scale=0.8]{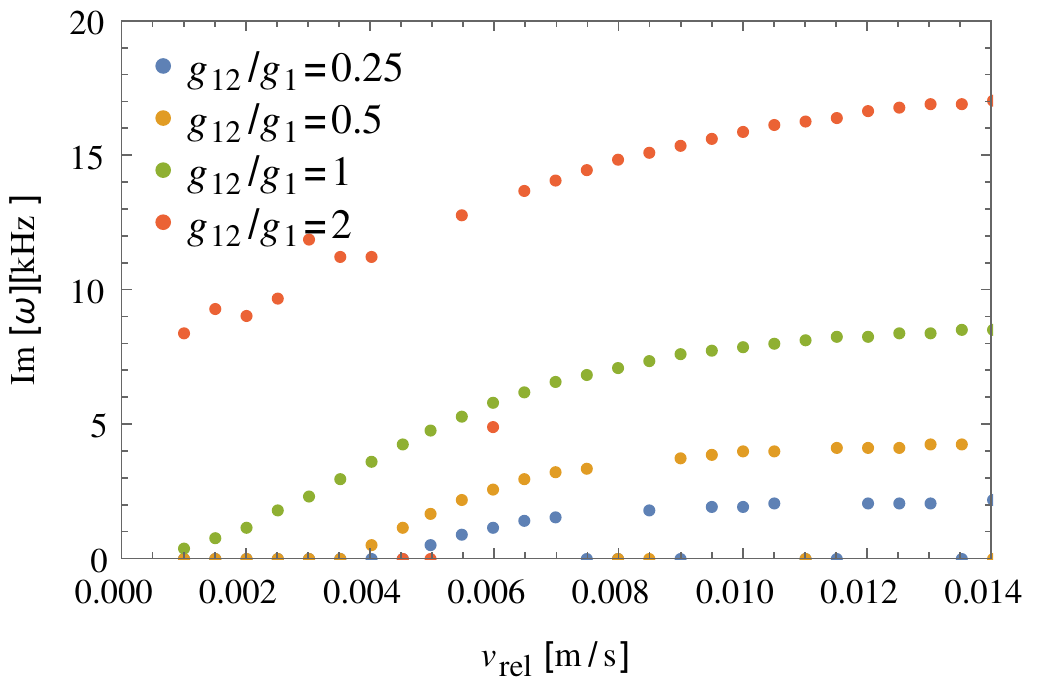}\\
(b) GPE\\
\includegraphics[scale=0.8]{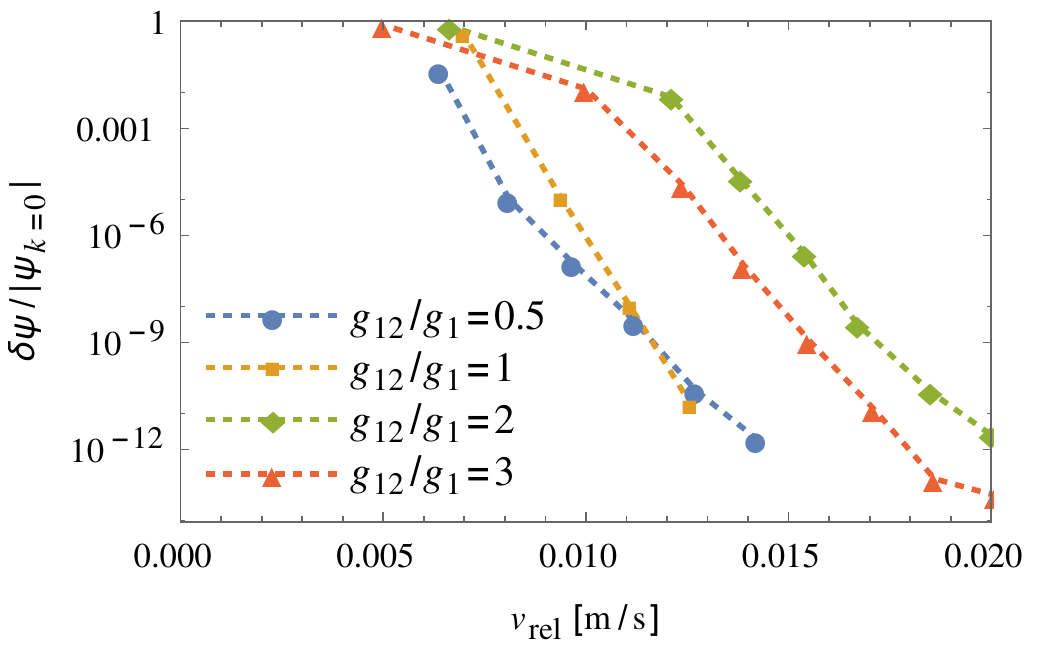}\\
\caption{Excitation amplitude as a function of relative velocity between the condensates. (a) In the Bogolibuov analysis, we computed the maximal value of ${\rm Im}[\omega]$. (b) In the GPE, we estimated the intensity of the excitation by subtracting the wavefunction at the collision time from the wavefunction at the initial state. See text for a discussion of the different behavior of the two curves.}\label{fig:Imax}
\end{figure}

We now discuss the excitation's intensity as a function of the relative velocity. For the Bogoliubov case, we considered the maximal value of the imaginary part of the excitation frequency, ${\rm max[Im(\omega)]}$. For the GPE, we computed the amplitude of the excitation peak during the collision (as explained above), $\delta\psi$. These quantities are shown in Fig.~\ref{fig:Imax}. As expected, in both cases, the intensity of the excitation is controlled by the interspecies interaction $g_{12}$, and tends to zero for $g_{12}\to 0$. 

The dependence of these two quantities as a function of the relative velocity is however very different. Concerning the Bogoliubov excitations (Fig.~\ref{fig:Imax}(b)), the imaginary part of the frequency grows with $v_{\rm rel}$. At small relative velocities, the excitations are suppressed due to the superfluid nature of the condensates: for relative velocities that are smaller than the critical velocity no excitation is generated (blue and yellow dots). For $g_{12}=g_1=g_2$, the system is at the miscible-to-immiscible phase transition and the critical velocity is exactly zero (green dots). For larger $g_{12}$ the system is unstable for any value of the critical velocity (red dots). 

In contrast, in the numerical solution of the GPE we find that intensity of the interactions {\it decreases} with the relative velocity. To understand this dependence, one needs to recall that dynamical instabilities are characterized by an exponential growth in time of pre-existing fluctuations. In the present case, the initial finite-wavevector fluctuations are determined by the intraspecies interactions. As shown in Fig.~\ref{collisions}, these fluctuations decay steeply as a function of the wavevector $k$. As a consequence, although the instability rate grows with $v_{\rm rel}$, the absolute intensity of the excitation actually decreases. This effect is further enhanced by the finite time of the collision: For large relative velocities the collision time is smaller and the excitation amplitude is further reduced. In actual experiments, the presence of thermal atoms is expected to enhance the initial fluctuations and to significantly increase the excitation intensity for large $v_{\rm rel}$.


\section{Conclusion}
In this article we studied the collision between two counterflowing condensates, using two complementary approaches. The first method consisted of an analytic expression for the Bogoliubov excitations of a translationally invariant system. The second approach involved the numerical solution of the Gross-Pitaevskii equation (GPE) and allowed us to study the effects of the trapping potential. For relative velocities that are larger than a critical velocity, the system develops a finite-wavevector instability, first predicted by Law \etal \cite{law2001critical}. We identified the origin of this instability as an exceptional point between two Bogoliubov modes, and confirmed its existence in finite systems. To guide future experiments, we characterized the instability through the excitation's wavevector and amplitude, as a function of the relative velocity and of the interspecies interaction. 

The Bogoliubov and GPE analyses delivered similar results and in particular confirmed that at large relative velocities, $v_{\rm rel}\gg c$, the wavevector of the instability is approximately given by the semiclassical expression $\hbar k = m v_{\rm rel}$. In this regime, the excitation amplitude decreases with $v_{\rm rel}$, due to the decrease of the initial fluctuations and to the shortening of the collision time. On the other hand, at small relative velocities $v_{\rm rel}\lesssim c$, the excitation amplitude is suppressed by the superfluid nature of the condensate, and tends to zero at the critical velocity of the system. Combining these two effects, we predict a non-monotonic dependence of the excitation amplitude on the relative velocity, with a maximum close to sum of the sound velocities of the two condensates. We hope that our theoretical calculations will guide and encourage experimentalists in the search for many-body exceptional points in colliding condensates. 

\emanuele{Our study shows that counterflowing condensates give rise to Bogoliubov modes with a complex frequency. Their dynamics can be effectively described by static, non-Hermitian Hamiltonians. Recent studies demonstrated that these models show interesting many-body effects, such as Kibble-Zurek mechanism \cite{yin2017kibble,zhai2018hybridized}, topological insulators \cite{lieu2018topological} and Majorana fermions \cite{san2016majorana,avila2018non}. An interesting question is whether these effects can be realized in the counterflow of condensates with richer (spin-orbit?) interactions.}

\emanuele{The main limitation of our calculation is the reliance on a ``mean-field'' approach, the GPE, where the operators $\psi_k$ and $\psi^\dagger_k$ are substituted by their expectation values. This approximation is justified for small momenta, whose occupation is very large \cite{kagan1997evolution} but is invalid for large momenta. Thus, to determine precisely the excitation amplitude for large relative velocities, it is necessary to take into account quantum corrections, associated with the minimal uncertainty between $\psi_k$ and $\psi^\dagger_k$. These effects can be treated using one of the known extensions of the GPE: the truncated Wigner method \cite{sinatra2001classical,sinatra2002truncated}, the stochastic GPE \cite{gardiner2002stochastic,gardiner2003stochastic}, and the time-dependent projected GPE \cite{davis2001simulations,blakie2005projected}. Alternatively, one can study this problem using general purpose quantum simulators, primarily the density matrix-renormalization group \cite{white1992density} and related methods, or the multiconfigurational time-dependent Hartree method for bosons \cite{streltsov2007role,alon2008multiconfigurational}. The proposed experiment offers a tunable testbed for the validity of these numerical methods.} 

{\bf Acknowledgment} This research is supported by the Israel Science Foundation Grant No. 1452/14. We acknowledge useful discussions with Frederic Chevy and Arnaud Courvoisier. We thank Yonathan Japha for sharing his code for the efficient numerical solution of the GPE.

\bibliography{researchbib}
\bibliographystyle{naturemag}

\end{document}